# Circular semi-quantum secret sharing using single particles

Chong-Qiang Ye, Tian-Yu Ye*

College of Information & Electronic Engineering, Zhejiang Gongshang University,
Hangzhou 310018, P.R.China

E-mail: happyyty@ailiyun.com

**Abstract:** Semi-quantum secret sharing (SQSS) is an important branch of semi-quantum cryptography, and differs from quantum secret sharing (QSS) in that not all parties are required to possess quantum capabilities. All previous SQSS protocols have three common features: (1) they adopt product states or entangled states as initial quantum resource; (2) the particles prepared by quantum party are transmitted in a tree-type way; and (3) they require the classical parties to possess the measurement capability. In this paper, two circular SQSS protocols with single particles are suggested, where the first one requires the classical parties to possess the measurement capability while the second one does not have this requirement. Compared with the previous SQSS protocols, the proposed SQSS protocols have some distinct features: (1) they adopt single particles rather than product states or entangled states as initial quantum resource; (2) the particles prepared by quantum party are transmitted in a circular way; and (3) the second protocol releases the classical parties from the measurement capability. The proposed SQSS protocols are robust against some famous attacks from an eavesdropper, such as the measure-resend attack, the intercept-resend attack and the entangle-measure attack, and are feasible with present quantum technologies in reality.

## 1 Introduction

Quantum cryptography provides a means of sending a secure message by the fundamental laws of quantum mechanics rather than the computational complexity of mathematical problems. Quantum key distribution (QKD) is one of the central problems in quantum cryptography. The first QKD protocol, i.e., the BB84 protocol, was proposed by Bennett and Brassard [1] in 1984. Afterward, many other QKD protocols [2-7] were proposed. Later, quantum information processing has been developed to cope with other cryptographic tasks, such as quantum secret sharing (QSS) [8-16], quantum authentication [17-19] and quantum secure direct communication (QSDC) [20-23] *etc*. In 1999, Hillery *et al.*[8] put forward the first QSS protocol by using three-particle entangled Greenberger-Horne-Zeilinger (GHZ) states. In 2003, Guo and Guo [11] proposed a novel QSS protocol with product states. In 2005, Deng *et al.*[12] presented an efficient QSS protocol with Einstein-Podolsky-Rosen pairs. In 2006, Deng *et al.*[13] put forward a circular QSS protocol based on single photons. Multiparty QSS protocols [14-15] have also been constructed based on single photons. The basis concept of QSS is that the boss distributes her secret among several agents using quantum technologies, and the secret can be recovered only when a sufficient number of agents collaborate together. A secure QSS protocol should be able to resist an attack from either an outside eavesdropper or an inside malicious participant.

In 2007, Boyer *et al.*[24-25] first proposed the novel concept named semi-quantum key distribution (SQKD) where Alice has full quantum capabilities but Bob is restricted to performing the following operations in quantum channel: (a) sending or returning the qubits without disturbance; (b) measuring the qubits in the fixed computational basis $\{|0\rangle,|1\rangle\}$; (c) preparing the (fresh) qubits in the fixed computational basis $\{|0\rangle,|1\rangle\}$; and (d) reordering the qubits (via different delay lines). According to the definition of the protocols in Refs.[24-25], the computational basis $\{|0\rangle,|1\rangle\}$ can be regarded as a classical basis, as it only refers to qubits $|0\rangle$ and $|1\rangle$ rather than any quantum superposition, and can be replaced with the classical notation $\{0,1\}$. It is of great interest to use as few quantum resource as possible to implement quantum cryptographic protocol. In 2009,

Zou and Qiu *et al.*[26] proposed other SQKD protocols with less quantum states. Thus, researchers have shown great enthusiasms on semi-quantum cryptography and have tried to apply the concept of semi-quantumness into different quantum cryptography tasks such as QKD, QSDC and QSS. As a result, many semi-quantum cryptography protocols, such as SQKD protocols [26-39], semi-quantum secure direct communication (SQSDC) protocols [40-42], semi-quantum secret sharing (SQSS) protocols [43-49], have been suggested.

In 2010, Li *et al.*[43] proposed two novel SQSS protocols by using GHZ-like states. In 2012, Wang *et al.*[44] presented an SQSS protocol by using two-particle entangled states. In 2013, Li and Qiu *et al.*[45] presented an SQSS protocol with two-particle product states; Lin *et al.*[46] pointed out that Li *et al.*'s two protocols [43] suffer from the intercept-resend attack and the Trojan horse attack from a dishonest agent, respectively, and suggested the corresponding improvements; Yang and Hwang [47] pointed out that desynchronizing the measurement operations among classical agents can enhance the efficiency of generating the shared key. In 2015, Xie *et al.*[48] proposed a novel SQSS protocol with GHZ-like states, where quantum Alice can share a specific bit string with classical Bob and classical Charlie instead of a random bit string. In 2016, Yin and Fu [49] proved that Xie *et al.*'s protocol [48] suffers from the intercept-resend attack from a dishonest party, and put forward an improved protocol accordingly. In summary, all previous SQSS protocols [43-49] have three common features: (1) they adopt entangled states or product states as initial quantum resource; (2) the particles prepared by quantum party are transmitted in a tree-type way, as depicted in Fig.1; and (3) they require the classical parties to possess the measurement capability.

Based on the above analysis, in this paper, we propose two circular SQSS protocols by using single particles as initial quantum resource, where the particles prepared by quantum party are transmitted in a circular way, as depicted in Fig.2. The first protocol requires the classical parties to possess the measurement capability while the second protocol does not have this requirement. Compared with the previous SQSS protocols, the proposed SQSS protocols have some distinct features: (1) they adopt single particles rather than product states or entangled states as initial quantum resource; (2) the particles prepared by quantum party are transmitted in a circular way; and (3) the second protocol releases the classical parties from the measurement capability.

The rest of this paper is organized as follows: in Sect.2, we propose two circular SQSS protocols with single particles and analyze their security in detail; and in Sect.3, discussion and conclusion are given.

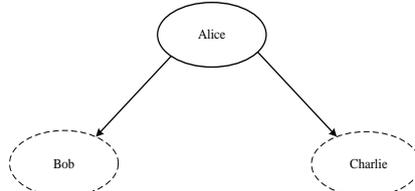

Fig.1  The particle transmission of the tree-type SQSS

In a tree-type SQSS, the shared key holder, quantum Alice, distributes different particles to classical Bob and classical Charlie, respectively. The solid line represents the quantum channel. The solid circle denotes the quantum party while the dotted circle represents the classical party.

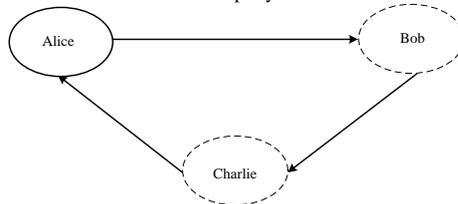

Fig.2  The particle transmission of the circular SQSS

In a circular SQSS, the particles from the shared key holder, Alice, run from Alice to Bob first, then from Bob to Charlie, and finally back to Alice.

## 2  Two circular SQSS protocols based on single particles

Suppose that there are three parties, Alice, Bob and Charlie. Alice has quantum capabilities while Bob and Charlie are restricted to only possessing classical capabilities. In this section, in order to accomplish the task of sharing Alice's key with Bob and Charlie, we construct two circular SQSS protocols by using single particles. The first one requires the classical parties to possess the measurement capability while the second one does not have this requirement. Throughout this paper, the measurement result of $|0\rangle$ ($|1\rangle$) stands for the classical bit 0 (1).

## 2.1 The circular SQSS protocol based on single particles requiring the measurement capability of classical parties

Before describing the protocol, we define two operations for both classical parties: (1) measuring the particle with $Z$ basis (i.e., $\{|0\rangle,|1\rangle\}$) and preparing a fresh one in the found state (referred to as MEASURE); and (2) reflecting the particle to the next receiver without disturbance (referred to as REFLECT).

**A. Protocol description**

The circular SQSS protocol based on single particles requiring the measurement capability of classical parties is composed of the following steps.

**Step 1:** Alice prepares $N+M$ ($M>N$) single particles, each of which is randomly in one of the four states $\{|0\rangle,|1\rangle,|+\rangle,|-\rangle\}$, and sends them to Bob.

**Step 2:** After receiving all particles from Alice, Bob makes two choices on them: (1) randomly MEASURE $N$ particles; and (2) REFLECT the remaining $M$ particles. Finally, Bob sends all of the $N+M$ particles in his hand to Charlie.

**Step 3:** After receiving all particles from Bob, Charlie also makes two choices on them: (1) randomly MEASURE $N$ particles; and (2) REFLECT the remaining $M$ particles. Finally, Charlie sends all of the $N+M$ particles in her hand to Alice.

**Step 4:** Alice confirms the receipt of $N+M$ particles from Charlie. After that, Bob and Charlie announce the positions of particles where they chose to MEASURE and where they chose to REFLECT, respectively.

**Step 5:** According to the choices of Bob and Charlie, Alice performs different operations on the received particles, as illustrated in Table 1. Note that Cases ① and ④ are only used to detect eavesdropping, while Cases ② and ③ can be used to share the secret key besides for eavesdropping detection.

Table 1  Actions implemented by Alice in four cases

| Case | Bob | Charlie | Alice |
|---|---|---|---|
| ① | MEASURE | MEASURE | ACTION 1 |
| ② | MEASURE | REFLECT | ACTION 1 |
| ③ | REFLECT | MEASURE | ACTION 1 |
| ④ | REFLECT | REFLECT | ACTION 2 |

ACTION 1: Alice measures the particles with $Z$ basis
ACTION 2: Alice measures the particles with her preparation basis

**Step 6:** Alice checks the error rates in Cases ① and ④ of Table 1, respectively. Concretely speaking, in Case ①, if there is no eavesdropping, all of the measurement results of Alice, Bob and Charlie on the corresponding particles will be the same. Alice let Bob and Charlie announce their measurement results respectively. Then Alice calculates the error rate of this case by comparing their three measurement results. If the error rate of this case exceeds the threshold value, the protocol will be terminated.

In Case ④, if there is no eavesdropping, Alice's measurement results will be the same as the corresponding states she prepared. Alice calculates the error rate of this case by comparing her measurement results with her corresponding prepared states. If the error rate in this case exceeds the threshold value, the protocol will be terminated.

**Step 7:** Alice checks the error rates in Cases ② and ③, respectively. Concretely speaking, in Case ②, if there is no eavesdropping, Alice's measurement results will be the same as Bob's measurement results on the corresponding particles. Alice randomly chooses a large enough subset of particles and announces their positions to Bob. Afterward, Bob publishes his measurement results on the corresponding particles to Alice. Then, Alice calculates the error rate of this case by comparing her measurement results with Bob's. If the error rate of this case exceeds the threshold value, the protocol will be terminated.

In Case ③, if there is no eavesdropping, Alice's measurement results will be the same as Charlie's measurement results on the corresponding particles. Alice randomly chooses a large enough subset of particles and announces their positions to Charlie. Afterward, Charlie publishes her measurement results on the corresponding particles to Alice. Then, Alice calculates the error rate of this case by comparing her measurement results with Charlie's. If the error rate of this case exceeds the threshold value, the protocol will be terminated.

**Step 8:** After the eavesdropping checks, Alice, Bob and Charlie can establish the key sharing relationship, i.e., $K_A = K_B \oplus K_C$. Here, $K_B$ is the bit string composed by the remaining measurement results from Bob in Case ②, $K_C$ is the bit string composed by the remaining measurement results from Charlie in Case ③, $K_A$ is Alice's shared bit string and $\oplus$ is the bitwise exclusive-OR operation. Note that Alice can extract $K_B$ and $K_C$ by implementing ACTION 1 on the corresponding particles in Cases ② and ③, respectively. Thus, only when Bob and Charlie collaborate together can they recover the key of Alice.

### B. Security analysis

Generally speaking, in a QSS protocol, the internal participants are more powerful than an external eavesdropper. It was pointed out in Ref.[9] that as to QSS, if the eavesdropping attacks from internal participants can be detected by security checks, the eavesdropping behavior from any eavesdropper (regardless of internal participants or an external eavesdropper) will be naturally discovered. Thus, in the following, we will only concentrate on analyzing the security of the proposed SQSS protocol against the dishonest parties. In the proposed SQSS protocol, the roles of Bob and Charlie are different, either of whom may be dishonest. Thus, we should demonstrate the security of the proposed SQSS protocol against dishonest Bob and dishonest Charlie, respectively.

**(1) Measure-resend attack**

*Suppose that Bob is the dishonest party.* In order to obtain Alice's shared key $K_A$, Bob needs to get Charlie's bit string generated in Case ③. In order to achieve this aim, Bob may try to launch the measure-resend attack. There are two kinds of measure-resend attack for Bob.

1) He measures all particles from Alice with $Z$ basis and sends the states he found to Charlie in Step 2. In Step 4, Bob pretends to announce $N$ positions where he chose to MEASURE and $M$ positions where he chose to REFLECT. After Charlie announces the positions of particles where she chose to MEASURE and where she chose to REFLECT in Step 4, Bob can easily get Charlie's bit string generated in Case ③. However, Bob's attack will be easily detected by the security check in Case ④ since Bob's measuring basis on the particles from Alice which both Bob and Charlie announced to REFLECT are not necessarily the same as Alice's preparation basis. For example, assume that Alice sends the particle $|+\rangle$ ($|-\rangle$) to Bob in Step 1. Bob measures $|+\rangle$ ($|-\rangle$) with $Z$ basis in Step 2. Without loss of generality, assume that after Bob's measurement, $|+\rangle$ ($|-\rangle$) is collapsed into the state $|0\rangle$. If both Bob and Charlie announce REFLECT on this particle, Alice will implement Action 2 on the state $|0\rangle$. In other words, Alice will measure $|0\rangle$ with $X$ basis (i.e., $\{|+\rangle, |-\rangle\}$). As a result, Alice will obtain $|+\rangle$ and $|-\rangle$ each with the probability of $50\%$. On the other hand, assume that Alice sends the particle $|0\rangle$ ($|1\rangle$) to Bob in Step 1. Bob measures $|0\rangle$ ($|1\rangle$) with $Z$ basis in Step 2 and gets the accurate state $|0\rangle$ ($|1\rangle$). If both Bob and Charlie announce REFLECT on this particle, Alice will implement Action 2 on the state $|0\rangle$ ($|1\rangle$). In other words, Alice will measure $|0\rangle$ ($|1\rangle$) with $Z$ basis. As a result, Alice will obtain the accurate state $|0\rangle$ ($|1\rangle$). To sum up, this kind of measure-resend attack from Bob will be detected by the security check in Case ④ with the probability of $25\%$.

2) Bob normally implements Step 2. When Charlie sends all particles to Alice in Step 3, Bob intercepts them, measures them with $Z$ basis and sends the states he found to Alice. In Step 4, Bob announces the positions of particles where he chose to MEASURE and where he chose to REFLECT according to what he did in Step 2. After Charlie announces the positions of particles where she chose to MEASURE and where she chose to REFLECT in Step 4, Bob can easily get Charlie's bit string generated in Case ③. However, this kind of measure-resend attack from Bob will be also detected by the security check in Case ④ with the probability of 25% since Bob's measuring basis on the particles from Charlie which both of them announced to REFLECT are not necessarily the same as Alice's preparation basis.

*Suppose that Charlie is the dishonest party.* In order to obtain Alice's shared key $K_A$, Charlie needs to get Bob's bit string generated in Case ②. In order to achieve this aim, Charlie may try to launch the measure-resend attack. There are two kinds of measure-resend attack for Charlie.

1) When Alice sends all particles to Bob in Step 1, Charlie intercepts them, measures them with $Z$ basis and sends the states she found to Bob. In Step 3, Charlie chooses to REFLECT all particles from Bob to Alice. In Step 4, Charlie pretends to announce $N$ positions where she chose to MEASURE and $M$ positions where she chose to REFLECT. After Bob announces the positions of particles where he chose to MEASURE and where he chose to REFLECT in Step 4, Charlie can easily get Bob's bit string generated in Case ②. However, this kind of measure-resend attack from Charlie will be detected by the security check in Case ④ with the probability of 25% since Charlie's measuring basis on the particles from Alice which both Bob and Charlie announced to REFLECT are not necessarily the same as Alice's preparation basis.

2) In Step 3, Charlie measures all particles from Bob with $Z$ basis and sends the states she found to Alice. In Step 4, Charlie pretends to announce $N$ positions where she chose to MEASURE and $M$ positions where she chose to REFLECT. After Bob announces the positions of particles where he chose to MEASURE and where he chose to REFLECT in Step 4, Charlie can get Bob's bit string generated in Case ②. However, this kind of measure-resend attack from Charlie will be detected by the security check in Case ④ with the probability of 25% since Charlie's measuring basis on the particles from Bob which both of them announced to REFLECT are not necessarily the same as Alice's preparation basis.

**(2) Intercept-resend attack**

*Suppose that Bob is the dishonest party.* In order to obtain Alice's shared key $K_A$, Bob needs to get Charlie's bit string generated in Case ③. In order to achieve this aim, Bob may try to launch the intercept-resend attack as follows. He implements Step 2 as usual with Alice's particles first. Then, he prepares $N+M$ fake single particles with $Z$ basis among which the $N$ single particles in the positions where he chose to MEASURE are in the states same as his measurement results. Afterward, he intercepts the particles Charlie sends to Alice and keeps them in his hand. Finally, he sends his fake single particles to Alice. After Charlie announces the positions of particles where she chose to MEASURE and where she chose to REFLECT in Step 4, Bob measures the corresponding particles from Charlie in his hand to get her bit string generated in Case ③. However, Bob's attack will be easily detected by either of the two security checks in Cases ③ and ④ since Bob's fake particles are not necessarily the same as the ones prepared by Alice.

*Suppose that Charlie is the dishonest party.* In order to obtain Alice's shared key $K_A$, Charlie needs to get Bob's bit string generated in Case ②. In order to achieve this aim, Charlie may try to launch the intercept-resend attack as follows. She prepares $N+M$ fake single particles with $Z$ basis first. Then, she intercepts the particles Alice sends to Bob and keeps them in her hand. Finally, she sends her fake single particles to Bob. Bob takes Charlie's fake single particles as normal particles and implements Step 2 as usual. Then, for Charlie, there are two choices of action.

1) Charlie implements Step 3 with the particles from Bob. After Bob announces the positions of particles where he chose to MEASURE and where he chose to REFLECT in Step 4, Charlie can easily get Bob's bit string generated in Case ② due to her preparation of fake single particles. However, Charlie's attack will be easily detected by either of the four security checks in Steps 6 and 7 since Charlie's fake particles are not necessarily the same as the ones prepared by Alice.

2) Charlie implements Step 3 with the particles from Alice instead of the ones from Bob. After Bob announces the positions of particles where he chose to MEASURE and where he chose to REFLECT in Step 4, Charlie measures the corresponding particles from Bob in her hand to get his bit string generated in Case ②. However, Charlie's attack will be easily detected by either of the two security checks in Cases ① and ② since Charlie's fake particles are not necessarily the same as the ones prepared by Alice. Moreover, in this situation, the key sharing relationship among Alice, Bob and Charlie, i.e., $K_A = K_B \oplus K_C$, cannot be correctly established, as Alice has no knowledge about $K_B$.

**(3) Entangle-measure attack**

The entangle-measure attack from an outside eavesdropper, Eve, is comprised of two unitaries: $U_E$ attacking particles as they go from Alice to Bob and $U_F$ attacking particles as they go back from Charlie to Alice, where $U_E$ and $U_F$ share a common probe space with the state $|\varepsilon\rangle$. As pointed out in Refs.[24-25], the shared probe allows Eve to make the attack on the returning particles depend on knowledge acquired by $U_E$ (if Eve does not take advantage of this fact, then the "shared probe" can simply be the composite system comprised of two independent probes). Any attack where Eve would make $U_F$ depend on a measurement made after applying $U_E$ can be implemented by $U_E$ and $U_F$ with controlled gates. Eve's entangle-measure attack within the implementation of the protocol is depicted in Fig.3.

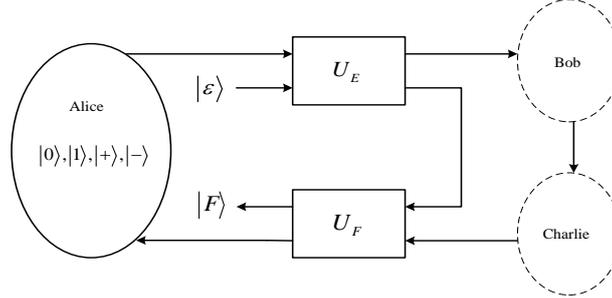

Fig.3　Eve's entangle-measure attack with two unitaries $U_E$ and $U_F$

***Theorem 1.*** *Suppose that Eve performs attack $(U_E, U_F)$ on the particles from Alice to Bob and back from Charlie to Alice. For this attack inducing no error in Steps 6 and 7, the final state of Eve's probe should be independent of Bob and Charlie's measurement results. As a result, Eve gets no information on Alice's shared key.*

***Proof.*** Before Eve's attack, the global state of the composite system composed by particles $A$ and $E$ is $|s\rangle_A |\varepsilon\rangle$, where $|s\rangle_A$ is the particle prepared by Alice which is randomly in one of the four states $\{|0\rangle, |1\rangle, |+\rangle, |-\rangle\}$. After Eve has performed $U_E$, the global state evolves into

$$\begin{cases} U_E(|s\rangle_A |\varepsilon\rangle) = |0\rangle_A |\varepsilon_{00}\rangle + |1\rangle_A |\varepsilon_{01}\rangle, & if\ |s\rangle_A = |0\rangle_A; \\ U_E(|s\rangle_A |\varepsilon\rangle) = |0\rangle_A |\varepsilon_{10}\rangle + |1\rangle_A |\varepsilon_{11}\rangle, & if\ |s\rangle_A = |1\rangle_A; \\ U_E(|s\rangle_A |\varepsilon\rangle) = |0\rangle_A |\varepsilon_{+0}\rangle + |1\rangle_A |\varepsilon_{+1}\rangle, & if\ |s\rangle_A = |+\rangle_A; \\ U_E(|s\rangle_A |\varepsilon\rangle) = |0\rangle_A |\varepsilon_{-0}\rangle + |1\rangle_A |\varepsilon_{-1}\rangle, & if\ |s\rangle_A = |-\rangle_A. \end{cases} \quad (1)$$

where $|\varepsilon_{00}\rangle$, $|\varepsilon_{01}\rangle$, $|\varepsilon_{10}\rangle$ and $|\varepsilon_{11}\rangle$ are un-normalized states of Eve's probe. Here, $|\varepsilon_{\pm x}\rangle = \frac{1}{\sqrt{2}}(|\varepsilon_{0x}\rangle \pm |\varepsilon_{1x}\rangle)$ and $x \in \{0,1\}$.

When Bob receives the particles from Alice, he chooses either to MEASURE or to REFLECT. When Charlie receives the particles from Bob, she also chooses either to MEASURE or to REFLECT. After that, Eve performs $U_F$ on the particles sent from Charlie back to Alice.

(i) Firstly, consider the case that both Bob and Charlie have chosen to MEASURE. As a result, the state of particles $A$ and $E$ after Bob and Charlie's operations will be

$$\begin{cases} |0\rangle_A |\varepsilon_{00}\rangle\ or\ |1\rangle_A |\varepsilon_{01}\rangle, & if\ |s\rangle_A = |0\rangle_A; \\ |0\rangle_A |\varepsilon_{10}\rangle\ or\ |1\rangle_A |\varepsilon_{11}\rangle, & if\ |s\rangle_A = |1\rangle_A; \\ |0\rangle_A |\varepsilon_{+0}\rangle\ or\ |1\rangle_A |\varepsilon_{+1}\rangle, & if\ |s\rangle_A = |+\rangle_A; \\ |0\rangle_A |\varepsilon_{-0}\rangle\ or\ |1\rangle_A |\varepsilon_{-1}\rangle, & if\ |s\rangle_A = |-\rangle_A. \end{cases} \quad (2)$$

Note that in this case, both Bob and Charlie own the measurement result of particle $A$ in their respective hands. For Eve not being detectable in the security check of Case ①, $U_F$ should meet the following relation:

$$U_F(|x\rangle_A|\varepsilon_{sx}\rangle) = |x\rangle_A|F_{sx}\rangle, \tag{3}$$

where $x \in \{0,1\}$ and $s \in \{0,1,+,-\}$. Eq.(3) means that $U_F$ cannot change the state of particle *A* after Bob and Charlie's operations. Otherwise, Eve will be detected with a non-zero probability in the security check of Case ①.

(ii) Secondly, consider the case that Bob has chosen to MEASURE and Charlie has chosen to REFLECT. As a result, the state of particles *A* and *E* after Bob and Charlie's operations can be also shown in Eq.(2). Note that in this case, only Bob owns the measurement result of particle *A* in his hand.

Assume that Bob's measurement result is $|0\rangle$. After Eve performs $U_F$ on the particle sent from Charlie back to Alice, due to Eq.(3), the state of particles *A* and *E* evolves into

$$\begin{cases} U_F(|0\rangle_A|\varepsilon_{00}\rangle) = |0\rangle_A|F_{00}\rangle, & if \quad |s\rangle_A = |0\rangle_A; \\ U_F(|0\rangle_A|\varepsilon_{10}\rangle) = |0\rangle_A|F_{10}\rangle, & if \quad |s\rangle_A = |1\rangle_A; \\ U_F(|0\rangle_A|\varepsilon_{+0}\rangle) = |0\rangle_A|F_{+0}\rangle, & if \quad |s\rangle_A = |+\rangle_A; \\ U_F(|0\rangle_A|\varepsilon_{-0}\rangle) = |0\rangle_A|F_{-0}\rangle, & if \quad |s\rangle_A = |-\rangle_A. \end{cases} \tag{4}$$

On the other hand, assume that Bob's measurement result is $|1\rangle$. After Eve performs $U_F$ on the particle sent from Charlie back to Alice, due to Eq.(3), the state of particles *A* and *E* evolves into

$$\begin{cases} U_F(|1\rangle_A|\varepsilon_{01}\rangle) = |1\rangle_A|F_{01}\rangle, & if \quad |s\rangle_A = |0\rangle_A; \\ U_F(|1\rangle_A|\varepsilon_{11}\rangle) = |1\rangle_A|F_{11}\rangle, & if \quad |s\rangle_A = |1\rangle_A; \\ U_F(|1\rangle_A|\varepsilon_{+1}\rangle) = |1\rangle_A|F_{+1}\rangle, & if \quad |s\rangle_A = |+\rangle_A; \\ U_F(|1\rangle_A|\varepsilon_{-1}\rangle) = |1\rangle_A|F_{-1}\rangle, & if \quad |s\rangle_A = |-\rangle_A. \end{cases} \tag{5}$$

Therefore, according to Eq.(4) and Eq.(5), no matter what Bob's measurement result is, Eve will be automatically not detected in the security check of Case ②.

(iii) Thirdly, consider the case that Bob has chosen to REFLECT and Charlie has chosen to MEASURE. As a result, the state of particles *A* and *E* after Bob and Charlie's operations can be also shown in Eq.(2). Note that in this case, only Charlie owns the measurement result of particle *A* in her hand.

Assume that Charlie's measurement result is $|0\rangle$. After Eve performs $U_F$ on the particle sent from Charlie back to Alice, due to Eq.(3), the state of particles *A* and *E* also evolves into Eq.(4).

On the other hand, assume that Charlie's measurement result is $|1\rangle$. After Eve performs $U_F$ on the particle sent from Charlie back to Alice, due to Eq.(3), the state of particles *A* and *E* also evolves into Eq.(5).

Therefore, according to Eq.(4) and Eq.(5), no matter what Charlie's measurement result is, Eve will be automatically not detected in the security check of Case ③.

(iv) Fourthly, consider the case that both Bob and Charlie have chosen to REFLECT. As a result, the state of particles *A* and *E* after Bob and Charlie's operations can be also shown in Eq.(1). Note that in this case, neither Bob nor Charlie own the measurement result of particle *A* in their respective hands.

After Eve performs $U_F$ on the particle sent from Charlie back to Alice, due to Eq.(3), the state of particles *A* and *E* evolves into

$$\begin{cases} U_F(|0\rangle_A|\varepsilon_{00}\rangle + |1\rangle_A|\varepsilon_{01}\rangle) = |0\rangle_A|F_{00}\rangle + |1\rangle_A|F_{01}\rangle, & if \quad |s\rangle_A = |0\rangle_A; \\ U_F(|0\rangle_A|\varepsilon_{10}\rangle + |1\rangle_A|\varepsilon_{11}\rangle) = |0\rangle_A|F_{10}\rangle + |1\rangle_A|F_{11}\rangle, & if \quad |s\rangle_A = |1\rangle_A; \\ U_F(|0\rangle_A|\varepsilon_{+0}\rangle + |1\rangle_A|\varepsilon_{+1}\rangle) = |0\rangle_A|F_{+0}\rangle + |1\rangle_A|F_{+1}\rangle = |+\rangle_A \frac{|F_{+0}\rangle + |F_{+1}\rangle}{\sqrt{2}} + |-\rangle_A \frac{|F_{+0}\rangle - |F_{+1}\rangle}{\sqrt{2}}, & if \quad |s\rangle_A = |+\rangle_A; \\ U_F(|0\rangle_A|\varepsilon_{-0}\rangle + |1\rangle_A|\varepsilon_{-1}\rangle) = |0\rangle_A|F_{-0}\rangle + |1\rangle_A|F_{-1}\rangle = |+\rangle_A \frac{|F_{-0}\rangle + |F_{-1}\rangle}{\sqrt{2}} + |-\rangle_A \frac{|F_{-0}\rangle - |F_{-1}\rangle}{\sqrt{2}}, & if \quad |s\rangle_A = |-\rangle_A. \end{cases} \tag{6}$$

For Eve not being detectable in the security check of Case ④, the following relations should be established:

$$|F_{01}\rangle = |F_{10}\rangle = 0, \tag{7}$$

$$|F_{+0}\rangle = |F_{+1}\rangle, \tag{8}$$

$$|F_{-0}\rangle = -|F_{-1}\rangle. \tag{9}$$

(v) According to Eq.(3), we have $U_F(|1\rangle_A|\varepsilon_{01}\rangle) = |1\rangle_A|F_{01}\rangle$ and $U_F(|0\rangle_A|\varepsilon_{10}\rangle) = |0\rangle_A|F_{10}\rangle$. As a result, it can be obtained from Eq.(7) that

$$|\varepsilon_{01}\rangle = |\varepsilon_{10}\rangle = 0. \tag{10}$$

Further, by inserting Eq.(10) into $|\varepsilon_{\pm x}\rangle = \frac{1}{\sqrt{2}}(|\varepsilon_{0x}\rangle \pm |\varepsilon_{1x}\rangle)$, we can have

$$|\varepsilon_{+0}\rangle = |\varepsilon_{-0}\rangle = \frac{1}{\sqrt{2}}|\varepsilon_{00}\rangle, \tag{11}$$

$$|\varepsilon_{+1}\rangle = -|\varepsilon_{-1}\rangle = \frac{1}{\sqrt{2}}|\varepsilon_{11}\rangle. \tag{12}$$

(vi) According to Eq.(3), we have $U_F(|0\rangle_A|\varepsilon_{+0}\rangle) = |0\rangle_A|F_{+0}\rangle$ and $U_F(|0\rangle_A|\varepsilon_{00}\rangle) = |0\rangle_A|F_{00}\rangle$. By combing Eq.(11) with these two equations, we can obtain

$$|F_{+0}\rangle = \frac{1}{\sqrt{2}}|F_{00}\rangle. \tag{13}$$

On the other hand, according to Eq.(3), we have $U_F(|1\rangle_A|\varepsilon_{+1}\rangle) = |1\rangle_A|F_{+1}\rangle$ and $U_F(|1\rangle_A|\varepsilon_{11}\rangle) = |1\rangle_A|F_{11}\rangle$. By combing Eq.(12) with these two equations, we can obtain

$$|F_{+1}\rangle = \frac{1}{\sqrt{2}}|F_{11}\rangle. \tag{14}$$

(vii) According to Eq.(3), we have $U_F(|0\rangle_A|\varepsilon_{-0}\rangle) = |0\rangle_A|F_{-0}\rangle$ and $U_F(|0\rangle_A|\varepsilon_{00}\rangle) = |0\rangle_A|F_{00}\rangle$. By combing Eq.(11) with these two equations, we can obtain

$$|F_{-0}\rangle = \frac{1}{\sqrt{2}}|F_{00}\rangle. \tag{15}$$

On the other hand, according to Eq.(3), we have $U_F(|1\rangle_A|\varepsilon_{-1}\rangle) = |1\rangle_A|F_{-1}\rangle$ and $U_F(|1\rangle_A|\varepsilon_{11}\rangle) = |1\rangle_A|F_{11}\rangle$. By combing Eq.(12) with these two equations, we can obtain

$$|F_{-1}\rangle = -\frac{1}{\sqrt{2}}|F_{11}\rangle. \tag{16}$$

(viii) After summing up Eqs.(8-9) and Eqs.(13-16), we have

$$|F_{+0}\rangle = |F_{+1}\rangle = |F_{-0}\rangle = -|F_{-1}\rangle = \frac{1}{\sqrt{2}}|F_{00}\rangle = \frac{1}{\sqrt{2}}|F_{11}\rangle. \tag{17}$$

According to Eq.(7) and Eq.(17), after ignoring the global factors, it can be concluded that for his attack inducing no error in Steps 6 and 7, the final state of Eve's probe should be independent of Bob and Charlie's measurement results. Thus, Eve gets no information on Alice's shared key.∎

In addition, if the dishonest Bob (Charlie) launches the entangle-measure attack shown in Fig.3, the following Lemma will be easily obtained in a similar way to Theorem 1.

**Lemma 1.** *Suppose that Bob (Charlie) performs attack $(U_E, U_F)$ on the particles from Alice to Bob and back from Charlie to Alice. For this attack inducing no error in Steps 6 and 7, the final state of Bob's (Charlie's) probe should be independent of Charlie's (Bob's) measurement results. As a result, Bob (Charlie) gets no information on Alice's shared key.*

## 2.2 The circular SQSS protocol based on single particles without requiring the measurement capability of classical parties

### A. Protocol description

The SQKD protocol in Ref.[37] releases the classical party from the classical basis measurement. Inspired by this protocol, in the following, we design a circular SQSS protocol based on single particles also releasing the classical parties from the classical basis measurement. The proposed circular SQSS protocol is composed of the following steps.

**Step 1:** Alice prepares $N$ single particles, each of which is randomly in one of the four states $\{|0\rangle, |1\rangle, |+\rangle, |-\rangle\}$, and sends them to Bob.

**Step 2:** After receiving all $N$ particles from Alice, Bob prepares $N$ new single particles with Z basis. Then, after reordering all $2N$ particles in his hand, Bob sends them together to Charlie.

**Step 3:** After receiving all $2N$ particles from Bob, Charlie prepares $N$ single particles with Z basis. Then, after reordering all $3N$ particles in her hand, Charlie sends them together to Alice. For convenience, the particles prepared by Alice, Bob and Charlie are denoted as CTRL particles, SIFT_B particles and SIFT_C particles, respectively.

**Step 4:** Alice publicly announces her receipt of the particles sent from Charlie. Then, Bob publishes the orders of the particles he sent to Charlie. In the same time, Charlie also publishes the orders of the particles she sent to Alice.

**Step 5:** Alice uses Z basis to measure SIFT_B particles and SIFT_C particles, and uses her preparation basis to measure CTRL particles. Subsequently, Alice checks the error rate on CTRL particles. If there is no eavesdropping, Alice's measurement results on CTRL particles will be the same as the states she prepared, as Bob and Charlie did nothing on them except the reordering operations. They will abort the protocol if the error rate on CTRL particles exceeds the threshold value.

**Step 6:** Alice randomly chooses a large enough subset of SIFT_B particles to be TEST particles. Then, Alice tells Bob the positions of the chosen SIFT_B particles. Afterward, Bob tells Alice the prepared states of these particles. Finally, Alice calculates the error rate by comparing her measurement results on these particles with Bob's prepared states. Alice will abort the protocol if the error rate exceeds the threshold value.

On the other hand, Alice randomly chooses a large enough subset of SIFT_C particles to be TEST particles. Then, Alice tells Charlie the positions of the chosen SIFT_C particles. Afterward, Charlie tells Alice the prepared states of these particles. Finally, Alice calculates the error rate by comparing her measurement results on these particles with Charlie's prepared states. Alice will abort the protocol if the error rate exceeds the threshold value.

**Step 7:** After the eavesdropping checks, Alice, Bob and Charlie can establish the key sharing relationship, i.e., $K_A = K_B \oplus K_C$. Here, $K_B$ is the bit string composed by Alice's measurement results on the remaining SIFT_B particles, $K_C$ is the bit string composed by Alice's measurement results on the remaining SIFT_C particles and $K_A$ is Alice's shared bit string. Thus, only when Bob and Charlie collaborate together can they recover the key of Alice.

### B. Security analysis

Same to our former protocol, in the following, we will only concentrate on analyzing the security of this proposed SQSS protocol against the dishonest parties. In this proposed SQSS protocol, the roles of Bob and Charlie are different, either of whom may be dishonest. Thus, we should demonstrate the security of this proposed SQSS protocol against dishonest Bob and dishonest Charlie, respectively.

**(1) Measure-resend attack**

*Suppose that Bob is the dishonest party.* In order to obtain Alice's shared key $K_A$, Bob needs to get Charlie's bit string. Thus, Bob may launch the measure-resend attack as follows. He normally implements Step 2 first. Then, in Step 3, he intercepts all particles from Charlie to Alice, measures them with $Z$ basis and sends the states he found to Alice. After Charlie publishes the orders of particles she sent to Alice in Step 4, Bob can easily get Charlie's bit string. Apparently, this kind of measure-resend attack from Bob can easily pass the security checks on SIFT_B particles and SIFT_C particles in Step 6, respectively. However, without knowing the genuine positions of CTRL particles when attacking, Bob will be detected inevitably by the security check on CTRL particles in Step 5. Concretely speaking, if Alice sends the particle $|0\rangle$ ($|1\rangle$) to Bob in Step 1, Bob's attack behavior will not change the state of this CTRL particle. In this case, no error will occur on this CTRL particle. On the other hand, if Alice sends the particle $|+\rangle$ ($|-\rangle$) to Bob in Step 1, after Bob intercepts it and measures it with $Z$ basis, this particle will be collapsed into $|0\rangle$ or $|1\rangle$ randomly. When Alice uses $X$ basis to check this CTRL particle in Step 5, she obtains the incorrect result with the probability of 50% in this case. To sum up, for each CTRL particle, Bob's measure-resend attack on it will be detected by the security check in Step 5 with the probability of 25%.

*Suppose that Charlie is the dishonest party.* In order to obtain Alice's shared key $K_A$, Charlie needs to get Bob's bit string. Thus, Charlie may launch the measure-resend attack as follows. After receiving all particles sent from Bob in Step 2, Charlie uses $Z$ basis to measure them and prepares the new particles in the states she found. Then she normally implements Step 3. After Bob publishes the orders of particles he sent to Charlie in Step 4, Charlie can easily get Bob's bit string. Apparently, this kind of measure-resend attack from Charlie can easily pass the security checks on SIFT_B particles and SIFT_C particles in Step 6, respectively. However, without knowing the genuine positions of CTRL particles when attacking, Charlie will be detected inevitably by the security check on CTRL particles in Step 5. Likewise, for each CTRL particle, Charlie's measure-resend attack on it will be detected by the security check in Step 5 with the probability of 25%.

**(2) Intercept-resend attack**

*Suppose that Bob is the dishonest party.* In order to obtain Alice's shared key $K_A$, Bob needs to get Charlie's bit string. In order to achieve this aim, Bob may try to launch the intercept-resend attack as follows. He implements Step 2 as usual first. Then, in Step 3, he intercepts all particles Charlie sends to Alice and keeps them in his hand. Afterward, he prepares $3N$ fake single particles with $Z$ basis and sends them to Alice. After Charlie publishes the orders of particles she sent to Alice in Step 4, Bob measures the corresponding particles from Charlie in his hand to get Charlie's bit string. However, Bob's attack on CTRL particles will be easily detected by Alice in Step 5, since Bob's corresponding fake particles are not necessarily the same as the ones prepared by Alice. Moreover, Bob's attack on SIFT_C particles will be easily detected by Alice in Step 6, since Bob's corresponding fake particles are not necessarily the same as the ones prepared by Charlie.

*Suppose that Charlie is the dishonest party.* In order to obtain Alice's shared key $K_A$, Charlie needs to get Bob's bit string. In order to achieve this aim, Charlie may try to launch the intercept-resend attack as follows. When Bob sends all particles to Charlie in Step 2, she keeps them in her hand. Then, in Step 3, she prepares $3N$ single particles with $Z$ basis and sends them to Alice. After Bob publishes the orders of the particles he sent to Charlie in Step 4, Charlie measures the corresponding particles from Bob in her hand to get Bob's bit string. However, Charlie's attack on CTRL particles will be easily detected by Alice in Step 5, since Charlie's corresponding fake particles are not necessarily the same as the ones prepared by Alice. Moreover, Charlie's attack on SIFT_B particles will be easily detected by Alice in Step 6, since Charlie's corresponding fake particles are not necessarily the same as the ones prepared by Bob.

**(3) Entangle-measure attack**

The particles Alice sends to Bob in Step 1 do not carry any useful information about Bob or Charlie's bit string. If an outside eavesdropper, Eve, is clever enough, she will not entangle her auxiliary particles with the particles Alice sends to Bob in Step 1 during their transmissions. The entangle-measure attack from Eve can be described as follows: the unitary $U_G$ attacking particles as they go from Bob to Charlie and the unitary $U_H$ attacking particles as they go back from Charlie to Alice. As for CTRL particles and SIFT_B particles, two unitaries, $U_G$ and $U_H$, share a common probe space with the state $|\varepsilon\rangle$. As pointed out in Refs.[24-25], the shared probe allows Eve to make the attack on the returning CTRL particles and SIFT_B particles depend on knowledge acquired by $U_G$ (if Eve does not take advantage of this fact, then the "shared probe" can simply be the composite system comprised of two independent probes). Any attack on CTRL particles and SIFT_B particles where Eve would make $U_H$ depend on a measurement made after applying $U_G$ can be implemented by $U_G$ and $U_H$ with controlled gates. As for SIFT_C particles, $U_G$ has nothing to do with them, and $U_H$ attacks them also with a probe space of the state $|\varepsilon\rangle$. Eve's entangle-measure attack within the implementation of the protocol is depicted in Fig.4.

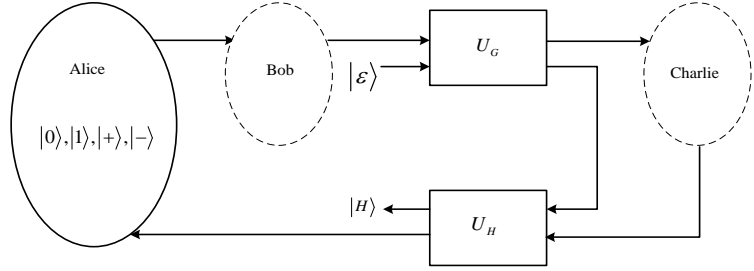

Fig.4  Eve's entangle-measure attack with two unitaries $U_G$ and $U_H$

**Theorem 2.** Suppose that Eve performs attack $(U_G, U_H)$ on the particles from Bob to Charlie and back from Charlie to Alice. For this attack inducing no error in Steps 5 and 6, the final state of Eve's probe should be independent of Bob and Charlie's prepared states. As a result, Eve gets no information on Alice's shared key.

**Proof.** Before Eve's attack, there are CTRL particles and SIFT_B particles in Bob's hand. Let $|s\rangle_A$ represent the CTRL particle and $|r\rangle_B$ denote the SIFT_B particle. Here, $|s\rangle_A \in \{|0\rangle, |1\rangle, |+\rangle, |-\rangle\}$ and $|r\rangle_B \in \{|0\rangle, |1\rangle\}$. After Eve has performed $U_G$, the composite system composed by particles $A$ and $G$ evolves into

$$\begin{cases} U_G(|s\rangle_A |\varepsilon\rangle) = |0\rangle_A |\varepsilon_{00}\rangle + |1\rangle_A |\varepsilon_{01}\rangle, & \text{if } |s\rangle_A = |0\rangle_A; \\ U_G(|s\rangle_A |\varepsilon\rangle) = |0\rangle_A |\varepsilon_{10}\rangle + |1\rangle_A |\varepsilon_{11}\rangle, & \text{if } |s\rangle_A = |1\rangle_A; \\ U_G(|s\rangle_A |\varepsilon\rangle) = |0\rangle_A |\varepsilon_{+0}\rangle + |1\rangle_A |\varepsilon_{+1}\rangle, & \text{if } |s\rangle_A = |+\rangle_A; \\ U_G(|s\rangle_A |\varepsilon\rangle) = |0\rangle_A |\varepsilon_{-0}\rangle + |1\rangle_A |\varepsilon_{-1}\rangle, & \text{if } |s\rangle_A = |-\rangle_A. \end{cases} \quad (18)$$

while the composite system composed by particles $B$ and $G$ evolves into

$$\begin{cases} U_G(|r\rangle_B |\varepsilon\rangle) = |0\rangle_B |\varepsilon_{00}\rangle + |1\rangle_B |\varepsilon_{01}\rangle, & \text{if } |r\rangle_B = |0\rangle_B; \\ U_G(|r\rangle_B |\varepsilon\rangle) = |0\rangle_B |\varepsilon_{10}\rangle + |1\rangle_B |\varepsilon_{11}\rangle, & \text{if } |r\rangle_B = |1\rangle_B. \end{cases} \quad (19)$$

Here, $|\varepsilon_{\pm x}\rangle = \frac{1}{\sqrt{2}}(|\varepsilon_{0x}\rangle \pm |\varepsilon_{1x}\rangle)$ and $x \in \{0,1\}$. After Eve has performed $U_H$, the composite systems composed by particles $A$ and $G$, particles $B$ and $G$, particles $C$ and $G$ evolve into

$$\begin{cases} U_H U_G(|s\rangle_A |\varepsilon\rangle) = U_H(|0\rangle_A |\varepsilon_{00}\rangle + |1\rangle_A |\varepsilon_{01}\rangle), & \text{if } |s\rangle_A = |0\rangle_A; \\ U_H U_G(|s\rangle_A |\varepsilon\rangle) = U_H(|0\rangle_A |\varepsilon_{10}\rangle + |1\rangle_A |\varepsilon_{11}\rangle), & \text{if } |s\rangle_A = |1\rangle_A; \\ U_H U_G(|s\rangle_A |\varepsilon\rangle) = U_H(|0\rangle_A |\varepsilon_{+0}\rangle + |1\rangle_A |\varepsilon_{+1}\rangle), & \text{if } |s\rangle_A = |+\rangle_A; \\ U_H U_G(|s\rangle_A |\varepsilon\rangle) = U_H(|0\rangle_A |\varepsilon_{-0}\rangle + |1\rangle_A |\varepsilon_{-1}\rangle), & \text{if } |s\rangle_A = |-\rangle_A. \end{cases} \quad (20)$$

$$\begin{cases} U_H U_G(|r\rangle_B |\varepsilon\rangle) = U_H(|0\rangle_B |\varepsilon_{00}\rangle + |1\rangle_B |\varepsilon_{01}\rangle), & \text{if } |r\rangle_B = |0\rangle_B; \\ U_H U_G(|r\rangle_B |\varepsilon\rangle) = U_H(|0\rangle_B |\varepsilon_{10}\rangle + |1\rangle_B |\varepsilon_{11}\rangle), & \text{if } |r\rangle_B = |1\rangle_B. \end{cases} \quad (21)$$

$$\begin{cases} U_H(|q\rangle_C |\varepsilon\rangle) = U_H(|0\rangle_C |\varepsilon\rangle), & \text{if } |q\rangle_C = |0\rangle_C; \\ U_H(|q\rangle_C |\varepsilon\rangle) = U_H(|1\rangle_C |\varepsilon\rangle), & \text{if } |q\rangle_C = |1\rangle_C. \end{cases} \quad (22)$$

respectively.

(i) For Eve not being detectable in the security check on CTRL particles in Step 5, the following relations should be established:

$$\begin{cases}
U_H U_G(|s\rangle_A|\varepsilon\rangle) = U_H(|0\rangle_A|\varepsilon_{00}\rangle+|1\rangle_A|\varepsilon_{01}\rangle) = |0\rangle_A|H_0\rangle, & \text{if } |s\rangle_A = |0\rangle_A; \\
U_H U_G(|s\rangle_A|\varepsilon\rangle) = U_H(|0\rangle_A|\varepsilon_{10}\rangle+|1\rangle_A|\varepsilon_{11}\rangle) = |1\rangle_A|H_1\rangle, & \text{if } |s\rangle_A = |1\rangle_A; \\
U_H U_G(|s\rangle_A|\varepsilon\rangle) = U_H(|0\rangle_A|\varepsilon_{+0}\rangle+|1\rangle_A|\varepsilon_{+1}\rangle) \\
\quad = \frac{1}{\sqrt{2}} U_H(|0\rangle_A|\varepsilon_{00}\rangle+|1\rangle_A|\varepsilon_{01}\rangle) + \frac{1}{\sqrt{2}} U_H(|0\rangle_A|\varepsilon_{10}\rangle+|1\rangle_A|\varepsilon_{11}\rangle) \\
\quad = \frac{1}{\sqrt{2}}(|0\rangle_A|H_0\rangle+|1\rangle_A|H_1\rangle) & \text{if } |s\rangle_A = |+\rangle_A; \\
\quad = \frac{1}{2}[|+\rangle_A(|H_0\rangle+|H_1\rangle)+|-\rangle_A(|H_0\rangle-|H_1\rangle)] \\
\quad = \frac{1}{2}|+\rangle_A(|H_0\rangle+|H_1\rangle), \\
U_H U_G(|s\rangle_A|\varepsilon\rangle) = U_H(|0\rangle_A|\varepsilon_{-0}\rangle+|1\rangle_A|\varepsilon_{-1}\rangle) \\
\quad = \frac{1}{\sqrt{2}} U_H(|0\rangle_A|\varepsilon_{00}\rangle+|1\rangle_A|\varepsilon_{01}\rangle) - \frac{1}{\sqrt{2}} U_H(|0\rangle_A|\varepsilon_{10}\rangle+|1\rangle_A|\varepsilon_{11}\rangle) \\
\quad = \frac{1}{\sqrt{2}}(|0\rangle_A|H_0\rangle-|1\rangle_A|H_1\rangle) & \text{if } |s\rangle_A = |-\rangle_A. \\
\quad = \frac{1}{2}[|+\rangle_A(|H_0\rangle-|H_1\rangle)+|-\rangle_A(|H_0\rangle+|H_1\rangle)] \\
\quad = \frac{1}{2}|-\rangle_A(|H_0\rangle+|H_1\rangle),
\end{cases} \quad (23)$$

From Eq.(23), it should hold that

$$|H_0\rangle = |H_1\rangle = |H\rangle. \quad (24)$$

After inserting Eq.(24) into Eq.(23), we have

$$\begin{cases}
U_H U_G(|s\rangle_A|\varepsilon\rangle) = |0\rangle_A|H\rangle, & \text{if } |s\rangle_A = |0\rangle_A; \\
U_H U_G(|s\rangle_A|\varepsilon\rangle) = |1\rangle_A|H\rangle, & \text{if } |s\rangle_A = |1\rangle_A; \\
U_H U_G(|s\rangle_A|\varepsilon\rangle) = |+\rangle_A|H\rangle, & \text{if } |s\rangle_A = |+\rangle_A; \\
U_H U_G(|s\rangle_A|\varepsilon\rangle) = |-\rangle_A|H\rangle, & \text{if } |s\rangle_A = |-\rangle_A.
\end{cases} \quad (25)$$

According to Eq.(25), for Eve not being detectable in the security check on CTRL particles in Step 5, Eve's final probe states should be independent from CTRL particles.

(ii) According to Eq.(23), the following relations automatically hold:

$$\begin{cases}
U_H U_G(|r\rangle_B|\varepsilon\rangle) = U_H(|0\rangle_B|\varepsilon_{00}\rangle+|1\rangle_B|\varepsilon_{01}\rangle) = |0\rangle_B|H_0\rangle, & \text{if } |r\rangle_B = |0\rangle_B; \\
U_H U_G(|r\rangle_B|\varepsilon\rangle) = U_H(|0\rangle_B|\varepsilon_{10}\rangle+|1\rangle_B|\varepsilon_{11}\rangle) = |1\rangle_B|H_1\rangle, & \text{if } |r\rangle_B = |1\rangle_B.
\end{cases} \quad (26)$$

After inserting Eq.(24) into Eq.(26), we have

$$\begin{cases}
U_H U_G(|r\rangle_B|\varepsilon\rangle) = U_H(|0\rangle_B|\varepsilon_{00}\rangle+|1\rangle_B|\varepsilon_{01}\rangle) = |0\rangle_B|H_0\rangle = |0\rangle_B|H\rangle, & \text{if } |r\rangle_B = |0\rangle_B; \\
U_H U_G(|r\rangle_B|\varepsilon\rangle) = U_H(|0\rangle_B|\varepsilon_{10}\rangle+|1\rangle_B|\varepsilon_{11}\rangle) = |1\rangle_B|H_1\rangle = |1\rangle_B|H\rangle, & \text{if } |r\rangle_B = |1\rangle_B.
\end{cases} \quad (27)$$

Eq.(27) means that Eve's final probe states are independent from SIFT_B particles, thus Eve cannot be detectable in the security check on SIFT_B particles in Step 6.

(iii) Combing Eq.(22) and Eq.(27), after ignoring the global factors, we have

$$\begin{cases}
U_H(|q\rangle_C|\varepsilon\rangle) = U_H(|0\rangle_C|\varepsilon\rangle) = |0\rangle_C|H\rangle, & \text{if } |q\rangle_C = |0\rangle_C; \\
U_H(|q\rangle_C|\varepsilon\rangle) = U_H(|1\rangle_C|\varepsilon\rangle) = |1\rangle_C|H\rangle, & \text{if } |q\rangle_C = |1\rangle_C.
\end{cases} \quad (28)$$

Eq.(28) means that Eve's final probe states are independent from SIFT_C particles, thus Eve cannot be detectable in the security check on SIFT_C particles in Step 6.

It can be concluded that for Eve inducing no error in Steps 5 and 6, the final state of Eve's probe should be independent of Bob and Charlie's prepared states. As a result, Eve gets no information on Alice's shared key.∎

In addition, if the dishonest Bob (Charlie) launches the entangle-measure attack shown in Fig.4, the following Lemma will be directly obtained from Theorem 2.

**Lemma 2.** Suppose that Bob (Charlie) performs attack $(U_G, U_H)$ on the particles from Bob to Charlie and back from Charlie to Alice. For this attack inducing no error in Steps 5 and 6, the final state of Bob's (Charlie's) probe should be independent of Charlie's (Bob's) prepared states. As a result, Bob (Charlie) gets no information on Alice's shared key.

## 3  Discussion and conclusion

In both of the proposed circular SQSS protocols, the particles prepared by quantum party are transmitted in a circular way. Thus, it is necessary to consider the Trojan horse attacks from an eavesdropper, including the invisible photon eavesdropping attack [50] and the delay-photon Trojan horse attack [51-52]. To prevent the invisible photon eavesdropping attack, the receiver can insert a filter in front of his devices to filter out the photon signal with an illegitimate wavelength before he deals with it [52-53]. To overcome the delay-photon Trojan horse attack, the receiver can use a photon number splitter (PNS) to split each sample quantum signal into two pieces and measure the signals after the PNS with proper measuring bases [52-53]. If the multiphoton rate is unreasonably high, this attack will be discovered.

We compare the proposed SQSS protocols with the SQSS protocols of Refs.[43-45,48]. Note that there are two SQSS protocols in Ref.[43], which are denoted as the protocol of Ref.[43]-A and the protocol of Ref.[43]-B, respectively. Moreover, the two proposed SQSS protocols of this paper are denoted as the proposed protocol A and the proposed protocol B, respectively. According to Table 2, it is clear that on the one hand, the proposed protocols just need single-particle states as initial quantum resource while the protocols of Refs.[43-44,48] and the protocol of Ref.[45] need quantum entangled states and two-particle product states as initial quantum resource, respectively; on the other hand, the proposed protocol B does not require the classical parties to possess the measurement capability while the protocols of Refs.[43-45,48] all have such requirement.

Table 2  Comparison between the proposed SQSS protocols and the SQSS protocols of Refs.[43-45,48]

|  | The protocol of Ref.[43]-A | The protocol of Ref.[43]-B | The protocol of Ref.[44] | The protocol of Ref.[45] | The protocol of Ref.[48] | The proposed protocol A | The proposed protocol B |
|---|---|---|---|---|---|---|---|
| Initial quantum resource | Three-particle entangled states | Three-particle entangled states | Two-particle entangled states | Two-particle product states | Three-particle entangled states | Single-particle states | Single-particle states |
| Particle transmission mode | Tree-type | Tree-type | Tree-type | Tree-type | Tree-type | Circular | Circular |
| Measurement capability of classical parties | Yes | Yes | Yes | Yes | Yes | Yes | No |

On the other hand, there always exist some differences for security proof between a tree-type SQSS protocol and a circular SQSS protocol. Without loss of generality, we take the comparison between the SQSS protocol of Ref.[45] and the proposed SQSS protocols for example to illustrate the differences.

In the proposed SQSS protocols, the two classical parties play the different roles due to the circular particle transmission mode. As a result, we have to demonstrate the security against dishonest Bob's attacks and dishonest Charlie's attacks, respectively, including the measure-resend attack and the intercept-resend attack. In the SQSS protocol of Ref.[45], it is not necessary to do so, as the two classical parties play the same roles, due to the tree-type particle transmission mode.

With respect to the entangle-measure attack from Eve, in the proposed protocol A, $U_E$ attacks particles as they go from Alice to Bob and $U_F$ attacks particles as they go back from Charlie to Alice; in the proposed protocol B, $U_G$ attacks particles as they go from Bob to Charlie and $U_H$ attacks particles as they go back from Charlie to Alice; and in the SQSS protocol of Ref.[45], $U$ attacks particles as they go from Alice to Bob and Charlie, and $V$ attacks particles as they go back from Bob and Charlie to Alice. Apparently, the above differences of the entangle- measure attack from Eve are caused by the different particle transmission modes between the proposed SQSS protocols and the SQSS protocol of Ref.[45].

In summary, in this paper, we propose two circular SQSS protocols by using single particles as initial quantum resource, where the particles prepared by quantum party are transmitted in a circular way. The first circular SQSS protocol requires the classical parties to possess the measurement capability while the second one does not have this requirement. Our analysis results show that the proposed SQSS protocols are robust against some famous attacks from an eavesdropper, such as the measure-resend attack, the intercept-resend attack and the entangle-measure attack. Compared with the previous SQSS protocols, the proposed SQSS protocols have some distinct features: (1) they adopt single particles rather than entangled states or product states as initial quantum resource; (2) the particles prepared by quantum party are transmitted in a circular way; and (3) the second protocol releases the classical parties from the measurement capability. In addition, the proposed SQSS protocols need the technologies of preparing, measuring and storing single particles, all of which are available in practice. Therefore, the proposed SQSS protocols are feasible with present quantum technologies in reality.

# Acknowledgments

Funding by the National Natural Science Foundation of China (Grant No.61402407) and the Natural Science Foundation of Zhejiang Province (Grant No.LY18F020007) is gratefully acknowledged.